\documentclass[journal,12pt]{IEEEtran}

\usepackage{spconf,amsmath,graphicx}
\usepackage{amssymb,color,epsfig,citesort}
\usepackage{algorithmic,algorithm}

\title{Multi-Stream Switching for Interactive Virtual Reality Video Streaming}
\name{Gene Cheung{\small $~^{\#}$},
Zhi Liu{\small $~^{\ast}$},
Zhiyou Ma{\small $~^{\$}$},
Jack Z. G. Tan{\small $~^{\$}$}}
\address{$^{\#}$\,National Institute of Informatics,
$^{\ast}$\,Waseda University,
$^{\$}$\,Kandao Technology}
\begin{document}
%
\maketitle
\begin{abstract}
Virtual reality (VR) video provides an immersive 360 viewing experience to a user wearing a head-mounted display: as the user rotates his head, correspondingly different fields-of-view (FoV) of the 360 video are rendered for observation.
Transmitting the entire 360 video in high quality over bandwidth-constrained networks from server to client for real-time playback is challenging.
In this paper we propose a multi-stream switching framework for VR video streaming: the server pre-encodes a set of VR video streams covering different view ranges that account for server-client round trip time (RTT) delay, and during streaming the server transmits and switches streams according to a user's detected head rotation angle.
For a given RTT, we formulate an optimization to seek multiple VR streams of different view ranges and the head-angle-to-stream mapping function simultaneously, in order to minimize the expected distortion subject to bandwidth and storage constraints.
We propose an alternating algorithm that, at each iteration, computes the optimal streams while keeping the mapping function fixed and vice versa.
Experiments show that for the same bandwidth, our multi-stream switching scheme outperforms a non-switching single-stream approach by up to 2.9dB in PSNR.
\end{abstract}
\vspace{-0.08in}
\section{Introduction}
\label{sec:intro}
\begin{figure}[h]
\centering
\includegraphics[width=9cm]{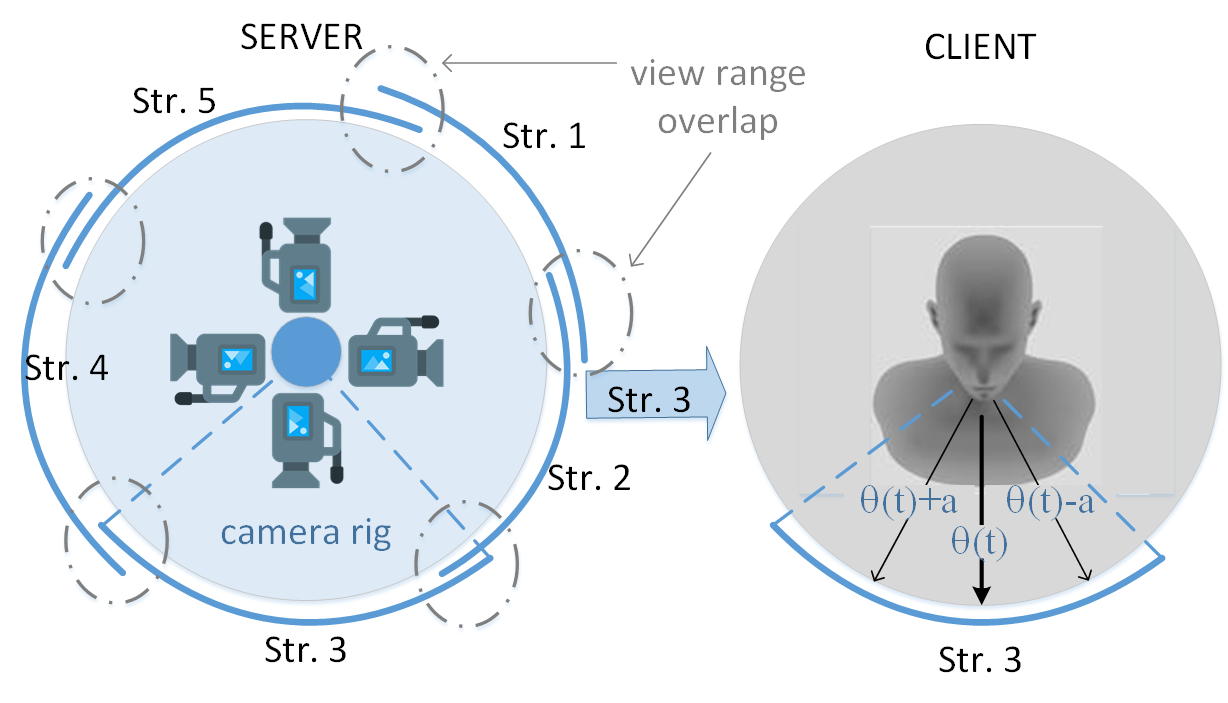}
\vspace{-0.4in}
\caption{\small{Interactive VR video streaming system using 5 pre-encoded streams with overlapping view ranges. Corresponding to user's head rotation angle $\theta(t)$ and FoV $[\theta(t)-a, \theta(t)+a]$ at time $t$, stream 3 is selected and transmitted.}}
\label{fig:vrStreaming}
\end{figure}

\vspace{-0.05in}
The advent of technologies for camera rigs, fisheye lenses and image-stitching algorithms \cite{jia08,zaragoza14} means that 360 \textit{virtual reality} (VR) video can now be readily generated.
A user equipped with a head-mounted display (HMD) such as Oculus Rift\footnote{https://www3.oculus.com/en-us/rift/} or HTC Vive\footnote{https://www.vive.com/jp/} can enjoy an immersive 360 viewing experience:
as the user rotates his head to the left or right, correspondingly different \textit{fields-of-view} (FoV) of the 360 VR video are rendered for observation.
See Fig.\;\ref{fig:vrStreaming} for an illustration.
It has been shown \cite{reichelt10} that such \textit{motion parallax} visual effect---changing FoVs according to user's head position and rotation angle---is the strongest cue for human's depth perception in a 3D scene, and VR video enables this effect for any head rotation angle from 0 to 360.

However, transmitting the entire 360 VR video in high quality over bandwidth-limited networks from a server to a client for real-time playback is challenging.
Leveraging on previous works in \textit{interactive multiview video streaming} (IMVS) \cite{cheung11tip,xiu12}, we propose a multi-stream switching framework for 360 VR video streaming.
The server pre-encodes a set of VR video streams, each covering a different view range of the original 360 video. During streaming, the server transmits and switches among the pre-encoded streams according to a user's detected head rotation angle.

By transmitting one video stream covering a limited view range at a time, the server can encode the stream at a higher quality than a single stream covering all 360 viewing angles for the same bandwidth constraint.
However, to minimize the adverse effect of interaction delay in motion parallax---even in the face of non-negligible server-to-client \textit{round trip time} (RTT) delay---each pre-encoded stream must cover a wide enough view range, so that a user's head with rotation angle starting in the view range center would not drift outside the view range in one RTT.
This implies that the coded streams tend to overlap in view ranges, resulting in representation redundancies and high storage cost.
Thus, \textit{multi-stream switching can enable higher visual quality, at the expense of an increase in storage cost due to streams' view range overlaps.}

Thus, the technical challenge is, for a given RTT, to design multiple VR streams of different view ranges and the head-angle-to-stream mapping function in order to minimize the expected distortion subject to bandwidth and storage constraints.
We mathematically formalize this optimization and propose an alternating algorithm that, at each iteration, computes the optimal VR streams while keeping the mapping function fixed and vice versa.
Experimental results show that for the same bandwidth constraint, our proposed multi-stream switching scheme outperforms a single-stream approach by up to $2.9$dB in PSNR.


\vspace{-0.08in}
\section{Related Work}
\label{sec:related}
\vspace{-0.05in}
Using an array of cameras to capture a 3D scene synchronously from slightly shifted viewpoints, IMVS systems \cite{cheung11tip,xiu12,maugey13,ren15,toni16} study how the captured multi-view videos can be pre-encoded into multiple streams. A receiving user can periodically request switches to neighboring camera views, and the server in response switches video streams with minimum discruption to the user's viewing experience.
To facilitate stream-switching, new frames like DSC frame \cite{mcheung09pcs} and merge frame (M-frame) \cite{dai16} were proposed.
Unlike IMVS \cite{cheung11tip,xiu12,maugey13,ren15,toni16}, we optimize the division of 360 VR video into multiple streams covering different view ranges given a constant RTT.
To the best of our knowledge, we are the first to study this problem for interactive VR video streaming formally. 

There are recent studies on VR video streaming. Assuming that the 3D scene can be represented by a 3D mesh, \cite{hosseini16,hosseini16_dc} proposed to first divide the mesh into 3D sub-meshes (tiles).
During streaming, a user communicates the desired tiles to the server using MPEG-DASH-SRD \cite{niamut16}, an extension of MPEG-DASH \cite{stockhammer11} to specify spatial relationships in media content.
Unlike \cite{hosseini16,hosseini16_dc}, we assume the input to our optimization is a 360 VR video, not 3D mesh. 
Further, we take the effect of RTT on interaction delay into account explicitly during optimization (to be detailed in Section \ref{sec:formulate}).

Assuming a camera rig with multiple cameras capturing a 360 view from different angles, \cite{sreedhar16} described a multiview video scheme that divides and codes captured camera views into two types: i) primary views at lower resolution that cover the entire 360 field-of-view, and ii) auxiliary views for the remaining camera views at high resolution.
The two video types are coded using multilayer extensions of HEVC. 
The receiver then performs image stitching to compose a 360 VR view. 
Instead, we assume 360 VR video is composed at the sender, and the challenge is to design multiple video streams covering different view ranges for interactive streaming.


\vspace{-0.08in}
\section{System Overview}
\label{sec:system}
\begin{figure}[h]
\centering
\includegraphics[width=8.5cm]{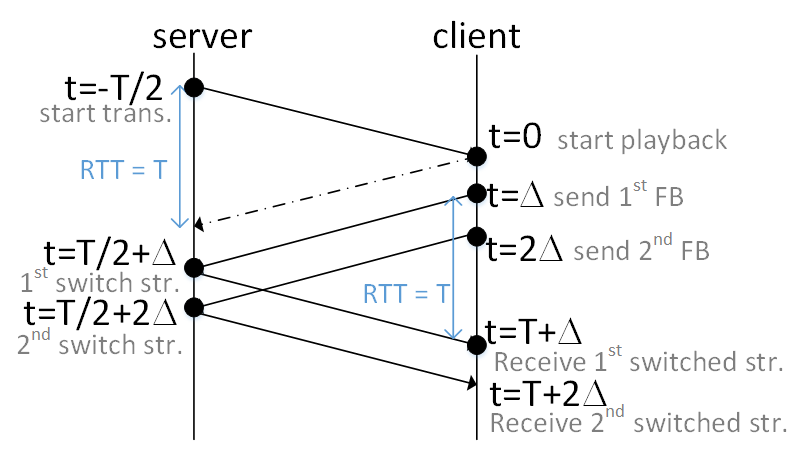}
\vspace{-0.2in}
\caption{\small{Interaction between server and client where RTT is $T$ and frame interval is $\Delta$. A switched stream arrives $T$ seconds after a feedback is sent.}}
\label{fig:RTT}
\end{figure}

\vspace{-0.05in}
We overview the operations of our multi-stream switching framework for a given RTT.
Denote by $T$ the RTT between server and client.
Denote by $\Delta$ the time interval between coded frames; $1/\Delta$ is the number of frames per second (fps).
For simplicity, assume for now that all frames are intra-coded, so that streams can be switched at any frame.
The server starts transmission of an initial video stream to the user at time $t=-T/2$, assuming the user begins at an initial head rotation angle $\theta(0)$.
At time $t=0$, the stream arrives at the client and playback begins.
At time $t=\Delta$, the client transmits the first feedback $\theta(\Delta)$ of the user's head rotation angle to the server.
This feedback $\theta(\Delta)$ arrives at the server at $t=T/2+\Delta$, and the server decides the new stream to transmit corresponding to $\theta(\Delta)$ using a \textit{mapping function} $f(\theta(\Delta))$.
This new stream arrives at the client at time $t=T+\Delta$, exactly $T$ seconds after feedback $\theta(\Delta)$ was generated.
Hence, \textit{the transmitted stream must accommodate the change in head rotation angle from $\theta(\Delta)$ to $\theta(T+\Delta)$.}
See Fig.\;\ref{fig:RTT} for an illustration.

Consider now the case when the VR streams are coded in Group-of-Pictures (GOP) of $H$ frames each.
This means that the frequency at which the server can switch streams is also every $H$ frames.
Compared to the previous case of intra-coded frames, each VR stream must now accommodate the change in head rotation angle in time interval $T + H \Delta$.

We next formulate the optimization problem to find the multiple VR streams and the mapping function $f(\,)$.


\vspace{-0.08in}
\section{Problem Formulation}
\label{sec:formulate}
\vspace{-0.05in}
\subsection{View Interaction Model}
\label{subsec:viewInteract}

\vspace{-0.05in}
We first define a \textit{view interaction model} that models a typical view selection process during 360 VR video observation.
Denote by $\theta[n]$ the \textit{central view angle} at which an observer is watching straight ahead at discrete time $n$. 
For convenience, we define the duration of a discrete time interval to be $\Delta$ (time interval between frames), and RTT in discrete instants to be $T_s = T/\Delta$.
We assume that $\theta[n] \in \{1, \ldots, K\}$ is also discrete, where $\theta[n] \, 2 \pi / K $ is angle in radians between $0$ and $2 \pi$.
We assume a one-hop Markov view transition model, where the probability of an observer's angle $\theta[n+1] = j$ given $\theta[n] = i$ is $p_{i,j}$.
Finally, we assume that the observer changes views only locally per instant, \textit{i.e.}, $p_{i,j} = 0 ~~\mbox{if} ~ |i-j| > v_{\max}$.

At any instant $n$, the observer has a FoV of size $1+2a \ll K$ that defines the angular span a human observes at a time.
Hence at time instant $n$, given central view angle $\theta[n]$, the observer's FoV is $\mathbf{R}[n] = [\theta[n] - a, \theta[n] + a]$.
It means that an observer will see visual distortion if the current video stream is not coded at high enough video quality in this range $\mathbf{R}[n]$.

\vspace{-0.05in}
\subsection{Expected Distortion}
\label{subsec:dist}

\vspace{-0.05in}
We define the expected distortion an observer sees in a 360 VR video as he naturally rotates his head.
We consider first the simple case when the GOP size is a single frame.
First, we compute the \textit{steady state probabilities} $\mathbf{q} \in \mathbb{R}^K$ assuming stationary view transition probabilities $p_{i,j}$ via the Perron-Frobenius Theorem\footnote{https://en.wikipedia.org/wiki/Perron\%E2\%80\%93Frobenius\_theorem}:
\begin{equation}
\mathbf{q} \, \mathbf{P}  = \mathbf{q}
\end{equation}
where $\mathbf{q}$ is the left eigenvector (row vector) corresponding to the eigenvalue $1$ for matrix $\mathbf{P}$.

Denote by $\mathbf{1}_{k}$ the canonical row vector of length $K$ with the only non-zero entry at position $k$ equals to $1$.
$T_s$ instants after an observer starts in central angle $k$, the angle distribution is $\mathbf{1}_k \mathbf{P}^{T_s}$.
Because an observer's FoV size is $1+2a$, we multiply $\mathbf{1}_k$ by a binary circulant matrix $\mathbf{C}_a \in \{0, 1\}^{K \times K}$ to account for FoV.
For example, $\mathbf{C}_1$ for $K=5$ is:
\begin{align}
\mathbf{C}_1 = \left[ \begin{array}{ccccc}
1 & 1 & 0 & 0 & 1 \\
1 & 1 & 1 & 0 & 0 \\
0 & 1 & 1 & 1 & 0 \\
0 & 0 & 1 & 1 & 1 \\
1 & 0 & 0 & 1 & 1
\end{array}
\right]
\end{align}

Suppose now that for central angle $k$, the server transmits stream $f(k)$ with distortion vector $\mathbf{d}_{f(k)}$, where $d_{f(k),l}$ is the distortion of angle $l$ in stream $f(k)$.
We can then write the expected distortion for this intra-coded streaming system as:
\begin{align}
D(\{\mathbf{d}_i\}, f(\,)) =
\sum_{k=1}^{K} q_k \; \mathbf{1}_{k} \mathbf{C}_a \mathbf{P}^{T_s} 
\mathbf{d}_{f(k)}
\label{eq:objectiveGOP1}
\end{align}
where the expected distortion $D$ depends on \textit{both} the distortion vectors $\mathbf{d}_i$ of different streams $i$ \textit{and} the mapping function $f(\,)$ from angles to streams.

If the 360 VR video streams are coded in GOP of $H$ frames each, then the stream-switching delay becomse $T_s + H$, and the distortion term for each $k$ needs to be computed for all $H$ frames:
\begin{align}
D(\{\mathbf{d}_i\}, f(\,)) =
\sum_{k=1}^{K} q_k \; 
\sum_{h=0}^{H-1} \mathbf{1}_{k} \mathbf{C}_a \mathbf{P}^{T_s+h} 
\mathbf{d}_{f(k)}
\label{eq:objectiveGOP2}
\end{align}

\vspace{-0.05in}
\subsection{Rate Constraints}

\vspace{-0.05in}
Given distortion vector $\mathbf{d}_i$ of stream $i$, we define the coding rate as $r(\mathbf{d}_i) = \sum_{k=1}^K  ~ g(d_{i,k})$, where $g(d_{i,k})$ is in turn defined as a clipped Laplacian function with parameter $\sigma$:
\begin{equation}
g(d) = U(d_{\max} - d) \exp \left( - \frac{|d|}{\sigma^2} \right)
\label{eq:codeRate}
\end{equation}
where $U(\,)$ is a step function; \textit{i.e.}, if $d \geq d_{\max}$, then rate $g(d)$ is $0$.
Parameter $\sigma$ can be chosen according to the 360 VR video characteristics.
Because distortion $d$ is non-negative, we can drop the absolute value operator in practice. 

Having defined $r(\mathbf{d}_i)$, we can define a storage constraint as follows.
Denote by $\mathcal{S}$ the set of pre-encoded video streams, by $Q$ the duration in time for the 360 video, and by $B$ the storage budget in bits.
We write the storage constraint as:
\begin{align}
\sum_{i \in \mathcal{S}} r(\mathbf{d}_i) \leq B/Q
\label{eq:storageConst}
\end{align}

We can similarly define a transmission constraint for a transmission budget $C$ in bps. 
Assuming a mapping function $f(\,)$ from angles to streams, we write:
\begin{align}
\sum_{k=1}^K q_k r(\mathbf{d}_{f(k)}) \leq C
\label{eq:transConst}
\end{align}

\vspace{-0.05in}
\subsection{Objective Function}
\label{subsec:obj}

\vspace{-0.05in}
Assuming $H=1$, collecting derived equations (\ref{eq:objectiveGOP1}), (\ref{eq:storageConst}) and (\ref{eq:transConst}), we write an unconstrained Lagrangian objective as:
\begin{equation}
\min_{\{\mathbf{d}_i\}, f(\,)}
\sum_{k=1}^{K} 
q_k \; \mathbf{1}_{k} \mathbf{C}_a \mathbf{P}^{T_s} 
\mathbf{d}_{f(k)} + 
\lambda \sum_{i \in \mathcal{S}} r(\mathbf{d}_i) +
\mu \sum_{k=1}^K q_k r(\mathbf{d}_{f(k)})
\label{eq:LagObj}
\end{equation}
where $\lambda$ and $\mu$ are chosen parameters so that the storage constraint (\ref{eq:storageConst}) and transmission constraint (\ref{eq:transConst}) are satisfied.

\vspace{-0.08in}
\section{Optimization Algorithm}
\label{sec:opt}
\vspace{-0.05in}
We take an alternating optimization approach, where we optimize variables $\{\mathbf{d}_i\}$ and $f(\,)$ one at a time while keeping the other fixed.
When $f(\,)$ is fixed, we take the derivative of the objective with respect to $d_{i,l}$ and set it to $0$:
\begin{align}
& \sum_{k | f(k) = i} q_k \left[ \mathbf{1}_{k} \mathbf{C}_a \mathbf{P}^{T_s} \right]_l +
\underbrace{\left( \lambda + \mu \sum_{k | f(k)=i} q_k \right)}_{\gamma}
\frac{\partial g(d_{i,l})}{\partial \, d_{i,l}} = 0 \nonumber \\
& - \frac{1}{\gamma} \sum_{k | f(k) = i} q_k \left[ \mathbf{1}_{k} \mathbf{C}_a \mathbf{P}^{T_s} \right]_l = 
\frac{\partial \exp \left( - \frac{d_{i,l}}{\sigma^2} \right)}{\partial \, d_{i,l}} \nonumber \\
& - \sigma^2 \log \left( \frac{\sigma^2}{\gamma} \sum_{k | f(k) = i} q_k \left[ \mathbf{1}_{k} \mathbf{C}_a \mathbf{P}^{T_s} \right]_l \right) = d_{i,l}^*
\label{eq:optdist}
\end{align}
where $[\,]_l$ denotes the $l$-th entry of a vector.

For intuition, we can check the boundary cases of (\ref{eq:optdist}) as follows.
If angle $l$ of stream $i$ is not observed (summation in the argument of $\log$ is 0), then the left side of (\ref{eq:optdist}) evaluates to $\infty$, so we can set $d^*_{i,k}$ to $d_{\max}$.
On the other hand, if angle $l$ is observed with high probability (summation in the argument of $\log$ is upper-bounded by $1$), assuming $\sigma^2/\gamma$ is also upper-bounded by $1$, then $d_{i,l}^*$ is lower-bounded by $0$.

When streams $\{d_i\}$ are fixed, we optimize $f(\,)$ simply as follows.
For each angle $k$, we identify a stream $i$ for $k$ with the minimum expected transmission cost in (\ref{eq:LagObj}). 

\vspace{-0.05in}
\subsection{Initialization}

\vspace{-0.05in}
For a given number $|\mathcal{S}|$ of target streams, we perform initialization as follows. 
We evenly distribute the central angles of $|\mathcal{S}|$ streams in $\{1, \ldots, K\}$.
For each stream $i$ with central angle $k$, we set distortion $d_{i,l}$ to a constant $d_1$ for angle $l$ where $|k-l| < T_s \, v_{\max}$; \textit{i.e.}, angle $l$ is reachable in $T_s$ transitions. Otherwise, $d_{i,l} = d_{\max}$. $d_1$ is then adjusted so that the transmission constraint is met for this stream.

The number of streams $|\mathcal{S}|$ is varied to find a locally optimal solution.

\vspace{-0.08in}
\section{Experiments}
\label{sec:results}

\vspace{-0.05in}
\subsection{Experimental Setup}

\vspace{-0.05in}
We use two 360 VR sequences captured by Kandao Technology\footnote{VR sequences will be made available at time of publication.}, \texttt{indoor concert} and \texttt{outdoor walking}, for our experiments.
Each video is 1 hour long at 30 fps.
FoV is assumed to be $90^{\circ}$, and $v_{\max}$ is $5^{\circ}$.
Video for one FOV has resolution $512 \times 512$.
Number of discrete view angles $K$ is 60, and RTT $T_s$ is $3$.
We use a linear function to model angle transition probabilities: $p_{i,j}$ linearly decreases with $|i-j|$, and the slope of decrease is steeper at $\pi/2$ and $3\pi/2$, resulting in higher steady state probabilities $q_k$ at these two angles.

As competitor we choose a non-switching scheme called \textit{static}, which always sends an encoded video covering the entire 360 angles.
For practical implementation, both our proposed scheme (called \textit{adaptive}) and \textit{static} use two QPs to encode each VR video stream; the two QPs are selected using \textit{Lloyd-Max quantizer} \cite{vq92} to approximate the theoretical Laplacian RD curve $r(d)$ shown in Fig.\;\ref{fig:result2}\,(a).
Test videos are first encoded at different QPs to generate empirical RD points, then the parameters of $r(d)$ are fitted.

\vspace{-0.05in}
\subsection{Experimental Results}

\begin{figure}[htb]
\begin{minipage}[b]{0.49\linewidth}
\centering
 \centerline{\includegraphics[width=4.7cm]{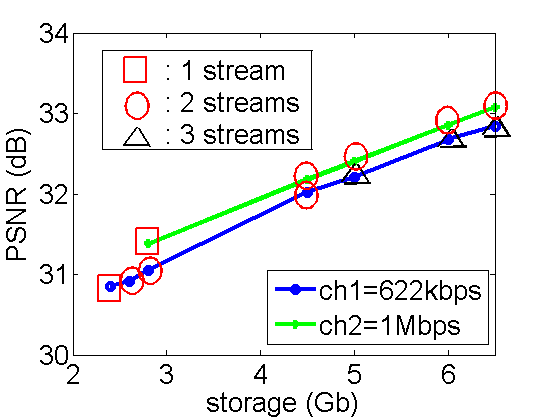}}
 \centerline{\footnotesize{(a) \texttt{indoor concert} }}\medskip
\end{minipage}
\hfill
\begin{minipage}[b]{0.49\linewidth}
  \centering
 \centerline{\includegraphics[width=4.7cm]{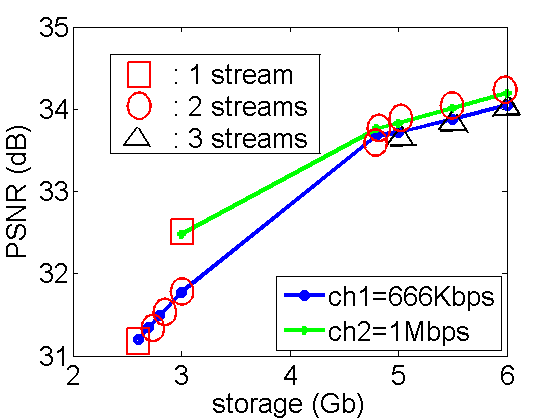}}
\centerline{\footnotesize{(b) \texttt{outdoor walking} }}\medskip
\end{minipage}
\vspace{-0.2in}
\caption{\small{PSNR versus storage for two competing schemes.}}%
\vspace{-0.1in}
\label{fig:result}
\end{figure}

\vspace{-0.05in}
We assume two different channel bandwidths are available.
We vary the available storage and show the tradeoff against visual quality (PSNR) in Fig.\;\ref{fig:result} for \texttt{indoor concert} and \texttt{outdoor walking}.
Weight parameters $\lambda$ and $\mu$ are tuned to satisfy bandwidth and storage constraints at each point.
Each data point in Fig.\;\ref{fig:result} is marked by a square, circle or triangle to denote the optimal number of streams generated: 1, 2 and 3, respectively.
\textit{static} uses 1 stream (squares), and \textit{adaptive} uses multiple streams (circles and triangles).

\begin{figure}[htb]
\begin{minipage}[b]{0.49\linewidth}
\centering
 \centerline{\includegraphics[width=4.6cm]{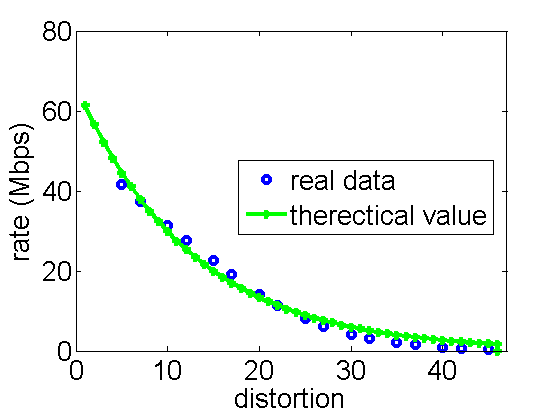}}
 \centerline{\footnotesize{(a) R-D curve }}\medskip
\end{minipage}
\hfill
\begin{minipage}[b]{0.49\linewidth}
  \centering
 \centerline{\includegraphics[width=4.6cm]{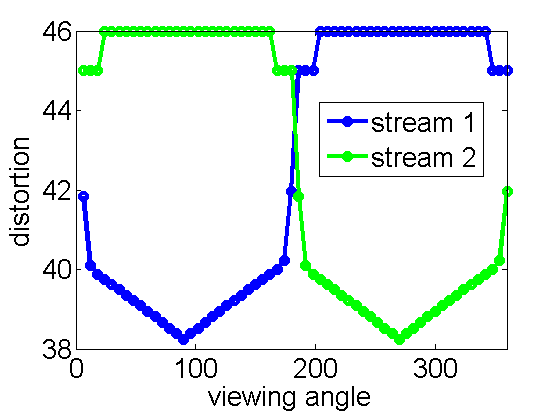}}
\centerline{\footnotesize{(b) distortion vs. view angle}}\medskip
\end{minipage}
\vspace{-0.2in}
\caption{\small{Illustration of R-D curve and streams' distortion vectors.}}%
\vspace{-0.1in}
\label{fig:result2}
\end{figure}

We observe that \textit{adaptive} outperforms \textit{static} for the two sequences---up to $2.9$dB in PSNR at the same bandwidth but using more storage.
For given channel bandwidth and storage, \textit{adaptive} selects the optimal number of streams and view range for each stream via optimization of distortion vectors $\mathbf{d}_i$.
Fig.\;\ref{fig:result2}\,(b) (distortion versus viewing angle) shows the optimized distortion vectors $\mathbf{d}_i$ for two streams when the storage is 5Gb and bandwidth is 1Mbps.
$d_{\max}=46$ in this case, and the corresponding angle range is not encoded because there is zero probability of being observed (given our view interaction model).
In contrast, viewing angles with high probabilities have low distortion values in $\mathbf{d}_i$.
We observe also that the two steams overlap, as discussed in the Introduction, to guarantee good visual quality when user's head rotates in one RTT.
Due to the low observe probability at the stream view range boundaries, the associated distortion values are relatively larger.

\vspace{-0.15in}

When storage is small, \textit{static} and \textit{adaptive} have the same performance for both $ch1$ and $ch2$.
As more storage becomes available, relative performance of \textit{adaptive} becomes better for both $ch1$ and $ch2$ when multiple streams are employed.
On the other hand, by sending only one stream always, \textit{static} cannot make use of extra storage to improve quality for a given channel bandwidth.




\vspace{-0.15in}


\vspace{-0.08in}
\section{Conclusion}
\label{sec:conclude}
\vspace{-0.05in}
Transmitting 360 VR video in high quality over bandwidth-limited networks is difficult.
In this paper, we pre-compute mulitple streams covering different overlapping view ranges at the server, and during streaming a single stream is selected corresponding to the user's tracked head rotation angle that minimizes the adverse effect of interaction delay.
We formulate an optimization to find the optimal streams and the head-angle-to-stream mapping function simultaneously, solved via an alternating algorithm.
Experimental results show that our multi-stream switching approach outperforms a single-stream approach by up to 2.9dB in PSNR.


\begin{small}
\bibliographystyle{IEEEbib}
\bibliography{ref2}
\end{small}

\end{document}